# Investigation about the electrochemical reduction in 3YSZ, related phase transition and consequences

Christian BECHTELER[1]

Richard I. TODD[1]

[1]*University of Oxford, Department of Materials, Parks Road, Oxford OX1 3PH, UK*

## Abstract

In this research the electrochemical reduction of 3YSZ was investigated in various atmospheres with different oxygen partial pressures under an electric field of 25 V/cm at an environmental temperature of 800 °C. At a certain oxygen partial pressure insufficient incorporation of oxygen in the sample led to electrochemical reduction of YSZ which shows two clearly distinguishable states. First, greying of the material without a significant change in properties was detected which then transitioned into a second stage where a fundamental phase transition in the material happened within seconds from tetragonal 3YSZ towards FCC rocksalt ZrO or ZrON, dependent on the atmosphere. This phase transition is accompanied by blackening of the material, sudden increase in electrical conductivity, current concentration, and an obvious change in Raman spectrum.

<u>Keywords:</u> electrochemical reduction, flash sintering, zirconium monoxide, zirconia, phase transformation

# Introduction

Yttria stabilised zirconia (YSZ) is the one of the most commonly used ceramics in high-temperature applications, gas sensors or as electrolyte material in solid oxide fuel cells (SOFCs) due to its pure oxygen ion conductivity, thermal and chemical stability, thermal shock resistance and good mechanical properties.

Even if FS caught attention again about a decade ago [1], a comparable approach was reported for heating zirconia about a century ago [2] and a practical application of current/voltage-supported heating of YSZ is the Nernst lamp was patented even before that in 1897 [3]. Flash sintering (FS), as it is used today, is a more sustainable field assisted sintering technique which enables the densification of ceramics at lower furnace temperatures and shorter time compared to conventional sintering [1], [4].

Rapid Joule heating as one important mechanism in FS is widely accepted [5]–[7]. However, some phenomena during electrical heating, like conducted during FS, can not solely be explained by Joule heating. Other potential fundamental mechanisms during electrical heating and FS in YSZ, as the most investigated material, but also other ceramics, remain partly unclear.

One potential mechanism is electrochemical reduction of YSZ, which leads to blackening of the material from the cathode side to the anode side in a DC field [8], [9] and a change from purely ionic at lower fields [10] to n-type semiconducting or electronic conduction [9], [11]–[13]. Potential blackening and a change in conduction mechanisms under FS conditions was also reported for other transition metal oxides such as ceria [14]. In an investigation by Biesuz et al. [9] the blackening of cubic Fluorite 8YSZ was investigated which showed that blackening can happen within a few seconds from cathode to anode in air and Ar. However, if the experiment was conducted in air, as it is usually the case, blackening disappeared because of reoxidation and it was not possible to characterise the blackened region. However, the investigation showed that blackening can happen and increases the materials conductivity in the voltage-controlled stage (I stage) of FS. This points towards the potential relevance and occurrence of electrochemical reduction under such conditions.

A more recent investigation by the authors [Bechteler et al.] showed that the electrochemical reduction of tetragonal YSZ leads to the formation of electronically conducting FCC rocksalt ZrO and ZrON, dependent on the atmosphere. However, in the previously published investigations [Bechteler et al. [15]–[17]], a possible phase transformation caused by electrochemical reduction was detected after a few minutes rather than seconds. In all publications [Bechteler et al. [15]–[17]], XRD experiments were used to detect new phases. Since, it is known that electrical properties can change from insulating to conducting if less than 0.25 wt.% of a conducting phase, with some kind of orientation or high aspect ratio, is introduced in an insulating matrix [18], the question arises if XRD is sensitive enough to detect an ongoing phase transition from the very beginning.

Thus, it remains unclear whether an electrochemically induced phase transformation can happen within seconds, and how could such a phase transition be recognized, otherwise its relevance to FS is rather low because FS happens within seconds not minutes. Furthermore, if a phase transformation happens in seconds, does it influence the materials behaviour under such conditions and could such a process be used to synthesise new materials in a sustainable way. Here, we investigate the possibility and the influence of electrochemical reduction in polycrystalline 3YSZ and how it affects its properties under a DC field and the possibility of a phase transformation happening within a few seconds.

## Experimental

Two sets of experiments were conducted starting at a furnace temperature of 800 °C. First, electrical loading of a pre-sintered fully dense 3YSZ (TZ-3YB-E, Tosoh, Japan) sample was conducted by stepwise (10 V every 30 s) increasing the applied DC field up to 50 V (= 25 V/cm) at. In this set of experiments a decrease in oxygen partial pressure under a constant field was realised by changing the atmosphere in the furnace from pure oxygen ($O_2$) to air to argon (Ar) or nitrogen ($N_2$). The change in atmosphere took a few seconds since the previous gas needs to be replaced to see the effect of the atmosphere change. The power dissipation in this set of experiments was stopped after 15 s after the change from air to Ar, or from air to $N_2$, or maintained during furnace cooldown. In another experiment the atmosphere was changed from $O_2$ to air to Ar and back to air to investigate the speed and influence of reoxidation. A power dissipation of 10 W was maintained for 3 min in Ar before the atmosphere was changed back to air.

In the second experiment, a green body made of 3YSZ was used and the electrical field was stepwise increased up to 350 V (=150 V/cm). A power dissipation of 25 W was maintained during FS in air including furnace cool down, which started after 5 min in the power-controlled stage. Turning off the power dissipation, as soon as the furnace temperature cooled down below 50 °C, led to quenching of the sample.

A DC-power supply (EA-PSI 9750-60; EA Elektro-Automatik GmbH & CO. KG, Germany) was used for applying the electrical field and recording the electrical parameters during the experiments. Platinum paste was used to improve the contact between the Pt-wire electrodes and the ceramic material.

Subsequent to the conducted experiments the samples were taken from the furnace and characterised by Raman spectroscopy (Reinshaw, 633 nm laser) within 5 minutes after taking samples out of the furnace to limit possible reoxidation. Some samples were cut to enable an investigation of the samples cross section. Furthermore, XRD characterisation of some of the samples was conducted with a scan speed of 0.04 °/s between 20 and 120 ° (Cu K$\alpha$ $\lambda$ = 1.5406 Å, Empyrean, Malvern Panalytical GmbH, Germany).

# Results

In the first set of experiments an increase in current (Figure 1a) and power dissipation (Figure 1b) can be detected during the voltage-controlled stage in pure $O_2$. The increase in current starts comparably slow before the increase becomes more rapidly and slows down again and plateaus at a current of 0.26 A after 550 s. This behaviour can be described by the current flow and Joule heating of the material which follows a NTC behaviour (Figure 1b), as expected for an ionic conductor.

After changing the atmosphere from pure $O_2$ to air (700 s) the current drops slightly to 0.24 A (Figure 1c). The sample up to this stage shows homogenous heating and no current concentration (Figure 2a). If the power dissipation is stopped at this point, and the sample is cooled down in inert atmosphere, the sample shows a homogeneous grey colour (Figure 2b).

After conducting the change in atmosphere from air to Ar after 1000 s (Figure 1d) the current starts increasing rapidly again after 1030 s. The increase in current at this point is significantly faster than it was detected at the beginning of the experiment in $O_2$. Furthermore, the current strongly concentrates in the sample and unlike before inhomogeneous sample heating is apparent (Figure 2c). If the current is stopped a few seconds after changing the atmosphere to Ar, at 1050 s in this case, the sample shows strong blackening after cool-down in inert atmosphere in the region where the current concentration happened. However, the cross section of the sample did not show blackening at this stage (Figure 2h) and the material on the cathode side, which was covered by Pt-paste, did not any greying or blackening. The electrical parameters for the same experiment in $N_2$, not Ar, are shown in the experimental results.

The detected blackening of the sample, where current concentrates, propagates along the surface of the pre-sintered sample (Figure 2f) with a speed of about 1.5 mm/s from the cathode (right) to the anode (left) of the sample, which agrees with previously reported findings in 8YSZ [9]. After 1053 s the transition from voltage to power control happens (Figure 1e) and a power spike, exceeding the defined limit of 25 W, can be detected for a few seconds (Figure 1f). At this point the blackened region propagated across the sample and connects cathode and anode.

If the current and the power dissipation is maintained for longer the current flow in the sample increases and sample heating shifts more towards the anode side of the sample (Figure 2e). After maintaining the power dissipation in the sample for a few hours, like described in a previous publication [Bechteler et al. 2023], the whole sample changed its colour towards a more metallic shining golden/brown colour (Figure 2f) and shows an electronic conduction and a metallic PTC behaviour rather than a NTC behaviour.

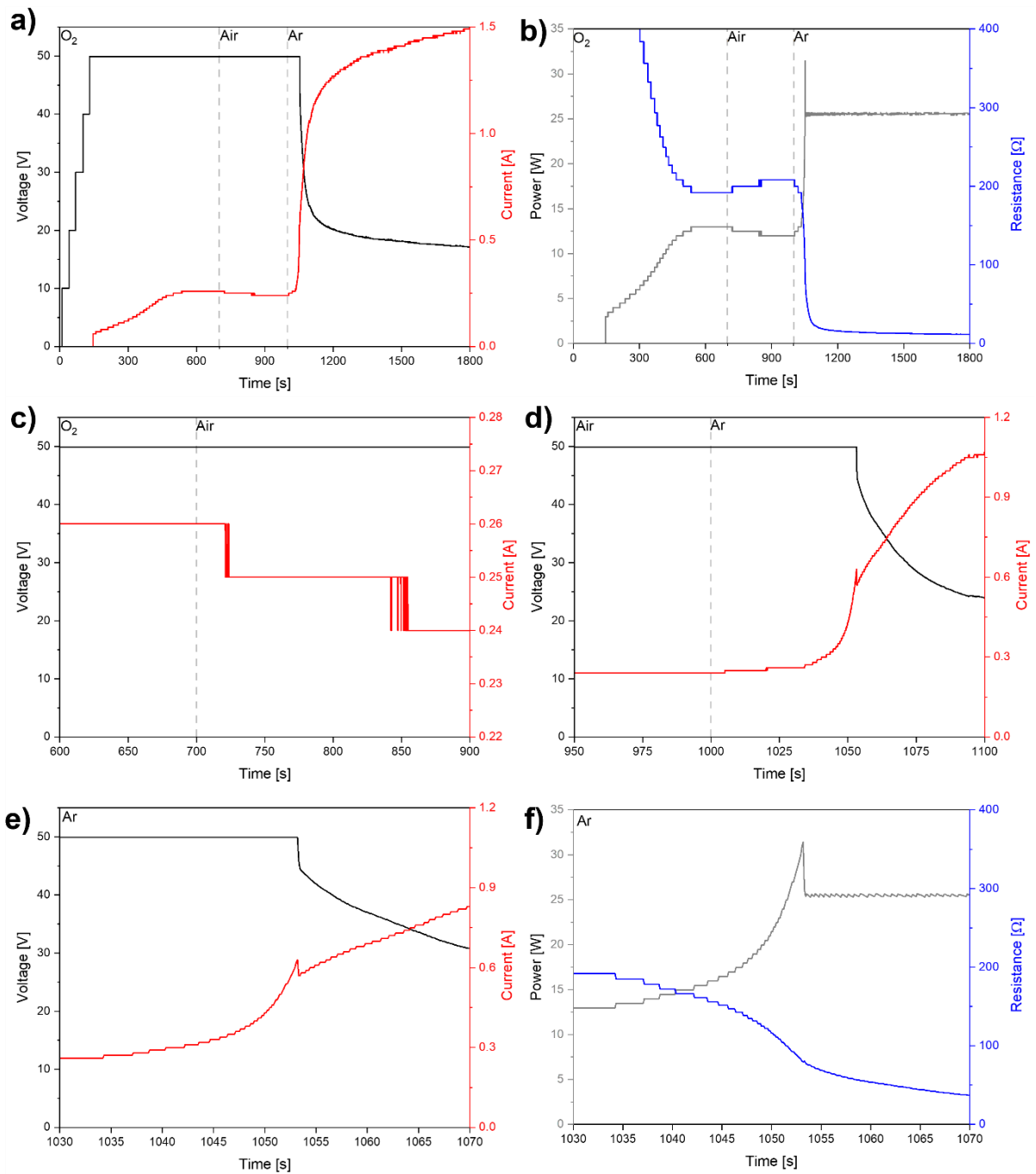

*Figure 1: Electrical parameters during FS of pre-sintered 3YSZ with decrease in oxygen partial pressure from pure oxygen to air to argon. The transition from oxygen to air happened after 700 s, the second change from air to argon after 1000 s. a) V and I during FS with various oxygen partial pressures; b) P and R during FS with various oxygen partial pressures; c) V and I during the change from oxygen to air; d) V and I during change from air to argon; e) V and I in argon atmosphere shortly after the atmospheric change; f) P and R in argon atmosphere shortly after the atmospheric change*

In the second experiment using a 3YSZ green body, which was flash sintered in air, the current concentrated in the centre of the sample as it is usually the case in FS. The power dissipation was maintained during furnace cool-down and subsequently to the experiment, the sample showed a white surface but a blackened region in the centre of the sample cross section which can be seen after cutting the sample (Figure 2g).

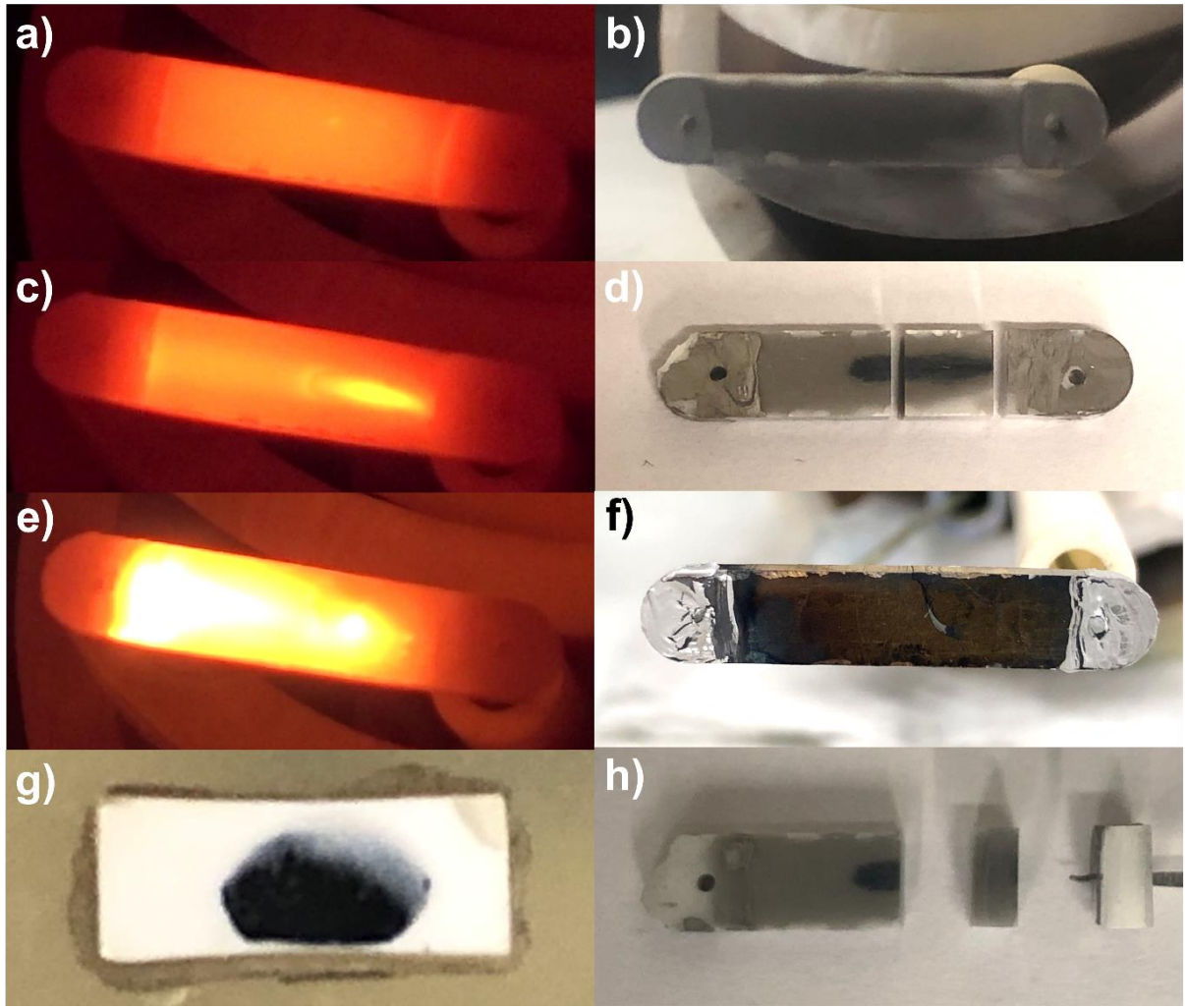

*Figure 2: Pre-sintered 3YSZ samples during and after applying an electrical field in various atmospheres. a) sample in and b) after electrical loading in air. c) and d) show sample shortly after the change from air to argon; e) and f) show sample after electrical loading for a few hours in argon maintained during furnace cool-down; g) shows cross section of a green 3YSZ sample after FS in air including power dissipation during furnace cooldown; h) shows cut sample and the cross section thereof as originally shown in d).*

After conducting the experiments, the different regions (white, grey, black) of the sample shown in Figure 2d) were characterised by Raman spectroscopy. Figure 3a shows the detected Raman spectra from the different regions of the sample. The white region shows the expected spectrum as it is known for tetragonal 3YSZ. In the grey region no significant change in the Raman spectrum was detected and the detected spectrum overlaps with the spectrum of the white region. In contrast to the white and the grey region, the black region in the sample shows a significantly different and more intense Raman spectrum.

The spectrum detected in the black region of the electrically loaded dense sample is comparable to the detected Raman spectrum in the black region of the cross section in the sample after FS in air (Figure 2g), which started as a green body. Furthermore, the comparison of the detected Raman spectra taken from pre-sintered samples which were electrically loaded for a few hours and are fully reduced to ZrO and ZrON, which show a comparable Raman spectrum to the black region in the previously described samples. The Raman spectrum detected in the white region of the FSed and quenched 3YSZ showed a different Raman spectrum to conventionally sintered 3YSZ, shown in Figure 3b), whereby the peaks of tetragonal YSZ are visible but the background signal is significantly stronger.

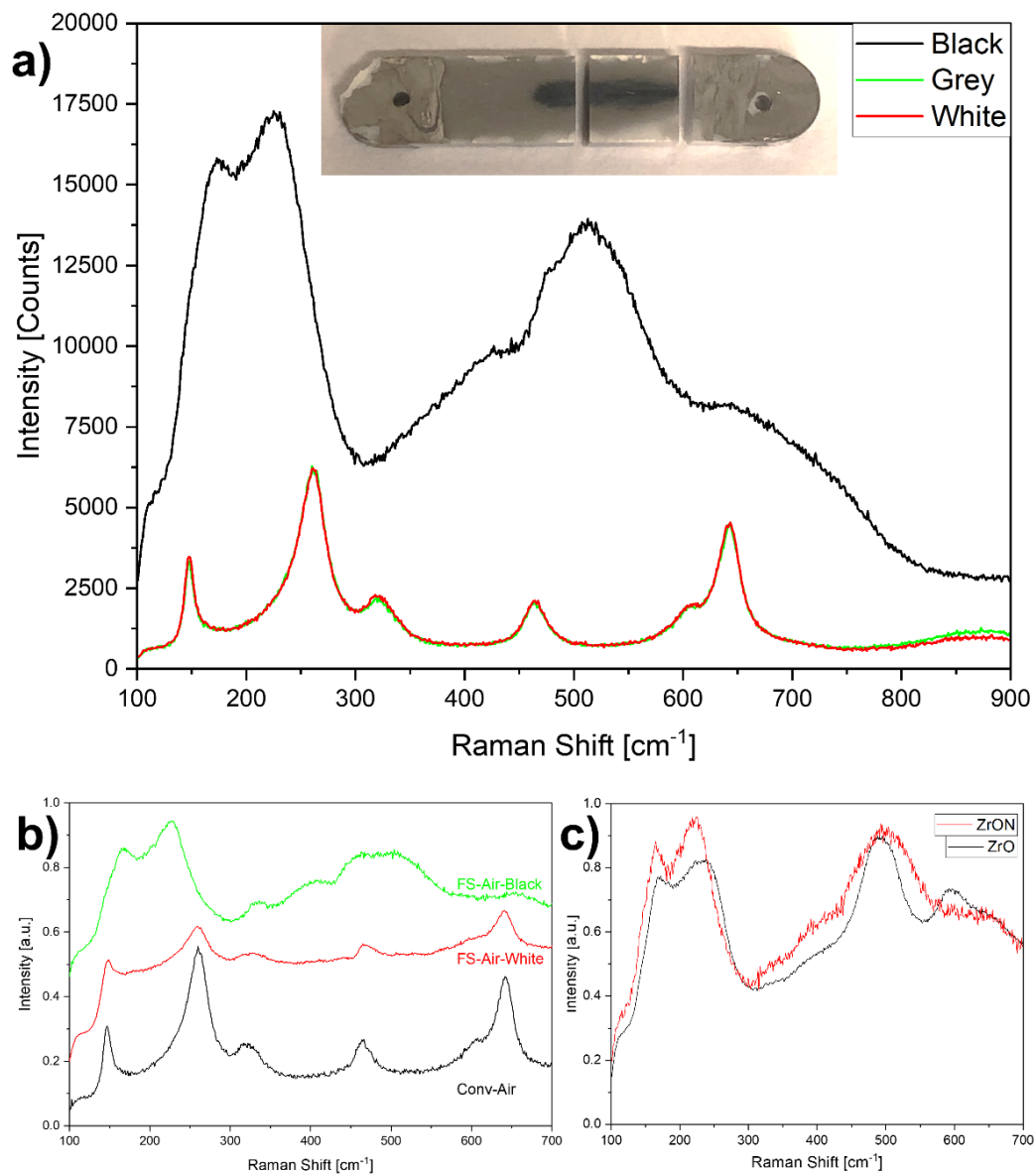

*Figure 3: Detected Raman spectra after applying an electrical field across the sample in atmospheres with decreasing oxygen partial pressure and stopping power dissipation shortly after change from air to argon. Black line shows spectrum detected in black region of the sample; green line represents grey area and red line represents white section in the sample; b) detected Raman spectra for flash sintered 3YSZ in the white region (red) in the black region (green) and the spectrum for conventionally sintered tetragonal 3YSZ; c) detected Raman spectra for ZrON and ZrO after fully reducing 3YSZ in $N_2$ or Ar, respectively.*

The XRD characterisation, shown in Figure 4, of the conventionally sintered and the right piece of the cut fraction of the partially reduced (Figure 2d) sample, did not show any unexpected peaks or significant peak changes in the material. Both samples showed the mainly tetragonal crystal structure and, unlike the fully reduced samples, no new phase or peaks were detected. Even the 102 peaks in tetragonal 3YSZ around 43 °, shown in more detail in Figure 4 b) is apparent in the conventionally sintered and partially reduced 3YSZ as well as the two characteristical {002} and {110} tetragonal peaks around 35 °. The fully reduced material shows a fundamentally changed structure including the formed FCC rocksalt ZrON phase with new peaks highlighted in form of grey stars on top of the peaks, which is discussed in detail in a previous publication [Bechteler et al.].

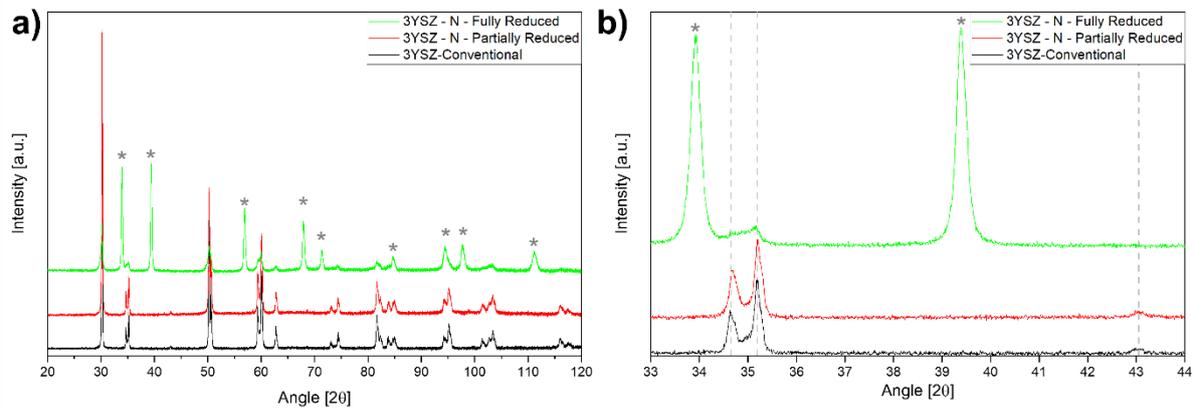

*Figure 4: a) Overall (20-120 °) and b) detailed (33-44 °) XRD spectra of conventionally sintered (black), partially reduced (red) and fully reduced (green) 3YSZ.*

Another question addressed in this investigation, is the possibility of reoxidation and how fast such a reoxidation could happen under such circumstances. After after dissipating 10 W for 3 min in Ar, reoxidation after changing the atmosphere back to air, happens within 2-3 s which is accompanied by a transition from the power-controlled stage to the voltage-controlled stage due to a significantly lower conductivity in the reoxidised state of the material. Such a reoxidation also happens as soon as the current is turned off in the furnace as it is usually done in other FS-experiments which makes the subsequent detection of a reoxidised phase almost impossible.

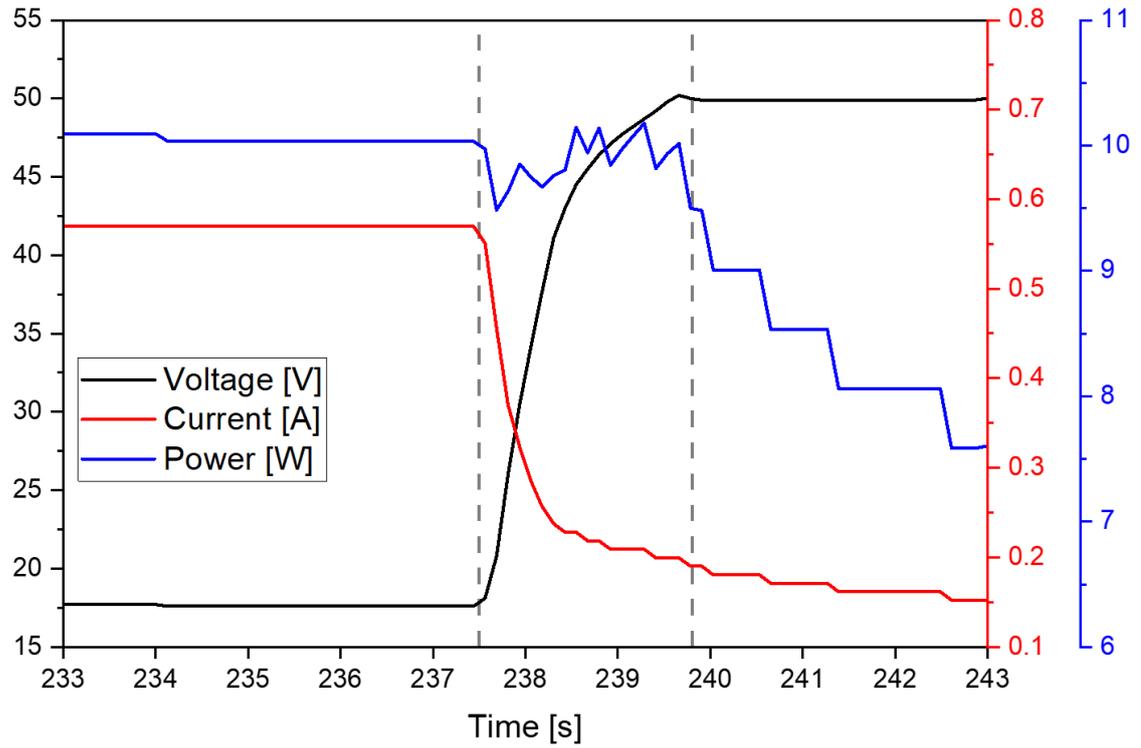

*Figure 5: Reoxidation of 3YSZ during the change from Ar to air atmosphere. Atmosphere was changed after 235 s Accompanied by a transition from power-controlled stage, ending at 237.5 s, to voltage-controlled stage, starting at approximately 240 s.*

## Discussion

The experiments conducted in this research showed that the electrochemical reduction of 3YSZ under the present circumstances, which are in the range of very moderate FS conditions, can be caused by a current flow through the sample and strongly depends on the oxygen partial pressure in the atmosphere. If the electrochemical reduction is below a certain value or threshold the fundamental structure of 3YSZ, according to Raman spectrum and XRD, remains mostly unchanged even if a greying of the material can happen. The electrically loaded sample in this stage shows no hotspot formation or inhomogeneous sample heating. It seems reasonable to describe partially reduced YSZ as $(Y)ZrO_{2-x}$, like it is usually done in the past, if the material is in this stage.

If the electrochemical reduction continues and the current flow in the material exceeds the necessary external oxygen ion supply, the material changes its fundamental structure and phase composition within seconds. If that is the case, the material changes from tetragonal YSZ to FCC rocksalt ZrO or ZrON, dependent on the atmosphere. This phase transition is accompanied by a change in colour from grey to black, change from ionic NTC to metallic PTC behaviour, the material becomes more conductive, and a current concentration and inhomogeneous heating occurs. In this stage the material should not be described as $(Y)ZrO_{2-x}$ because it changed its properties and fundamental structure in this stage.

Since blackening in YSZ as well as other transition metal oxides was reported previously [8], [9], [14], it needs to be questioned if in these cases a potential phase transition happened. The relevance of oxygen partial pressure on the behaviour of YSZ under FS conditions was reported previously [19] which showed that thermal runaway at higher oxygen partial pressures under constant field happened at higher furnace temperatures. This agrees on our results about the electrically loaded sample which showed a significantly slower increase in current and power dissipation in pure oxygen compared to the point where the electrochemical reduction in Ar or $N_2$ happened. These results highlight the importance of a sufficient $O_2$ supply during electrical heating, particularly under DC fields, if electrochemical reduction needs to be prohibited from the very beginning.

If electrochemical reduction occurs, a phase transition from an early stage on can be detected by Raman spectroscopy, whereas XRD did not show significant changes at an early stage. Since the Raman spectrum of the fully reduced sample, which contains significant amounts of ZrO(N), shows a comparable Raman spectrum to the one detected after partial reduction within a few seconds after changing atmosphere from air to Ar or $N_2$, it can be concluded that the phase transition happens within seconds at a moderate electrical field of 25 V/cm.

The two potential reasons for the missing change in XRD are the amount of transformed phase was below the detection limit of our XRD experiment and the formed crystallite size is too small to be detected. This highlights that even if only a small fraction of material gets sufficiently reduced the overall properties change significantly. The small crystallite size being another reason for the missing XRD response, is supported by the previously reported decrease in FWHM of the XRD spectrum detected for ZrO and ZrON after electrical loading YSZ with increasing power dissipation for a few hours [Bechteler et al. 2023].

According to the literature the FCC rocksalt structures do not allow for first-order Raman scattering. However, it is well-known that even in stoichiometric FCC rocksalt structures, including transition metals such as Zr[20], [21], Ti [22] or Hf[23], show sublattice defects like vacancies or antisite defects on the cationic and anionic sites, which cause and introduce the detected first-order Raman modes. It seems quite likely that in the detected FCC rocksalt ZrO(N) some degree of disordering is common since the materials in this research were quenched by a few hundred degrees after turning off the current flow and power dissipation. The potential disordering and defect structure in the FCC rocksalt ZrO(N) could also be one reason why, unlike suggested by other researchers [24], the detected structures are FCC rocksalt rather than hexagonal. A more detailed investigation of the disordering and more fundamental structure and properties of the FCC rocksalt ZrO and ZrON phases is ongoing.

Another important effect, happening in seconds, is reoxidation of the electrochemically reduced phases. In the case of pure zirconia, fully oxidised cubic Fluorite $ZrO_2$ structure shows lattice parameters around 0.515 nm which results in a unit cell volume of approximately 0.137

nm$^3$. FCC rocksalt ZrO shows a lattice parameter of 0.458 nm and a unit cell volume of 0.096 nm$^3$, which is about 70 % of the unit cell of Fluorite ZrO$^2$. The reoxidation of ZrO(N) to ZrO$^2$ leads to a significant expansion of the material which can lead to cracking of the material, but also, closure of pores. The phenomenon of pore closure by reoxidation is known in other fields such as thermal spry coating [25]. If the reoxidation process under certain FS conditions can be another reason, beside Joule heating, for the rapid densification, remains to be seen but seems possible if FS is conducted in a regime where reduction and phase transitioning happens.

Since this research preliminarily showed that an electrochemically induced phase transition in tetragonal YSZ towards FCC rocksalt ZrO(N) can happen within seconds, the possibility of such an effect occurring in other materials needs to be explored in the future. This is necessary to understand the behaviour and influence of electrical fields on the behaviour of materials, particularly transition metals and their oxides, under electrical fields. Furthermore, the investigated process provides a sustainable possibility to synthesise new materials at comparably low furnace temperature which can be modified by the present atmosphere and the applied conditions.

# Acknowledgements
This work was supported by the Engineering and Physical Sciences Research Council [EP/T517811/1].

# Supplementary Information

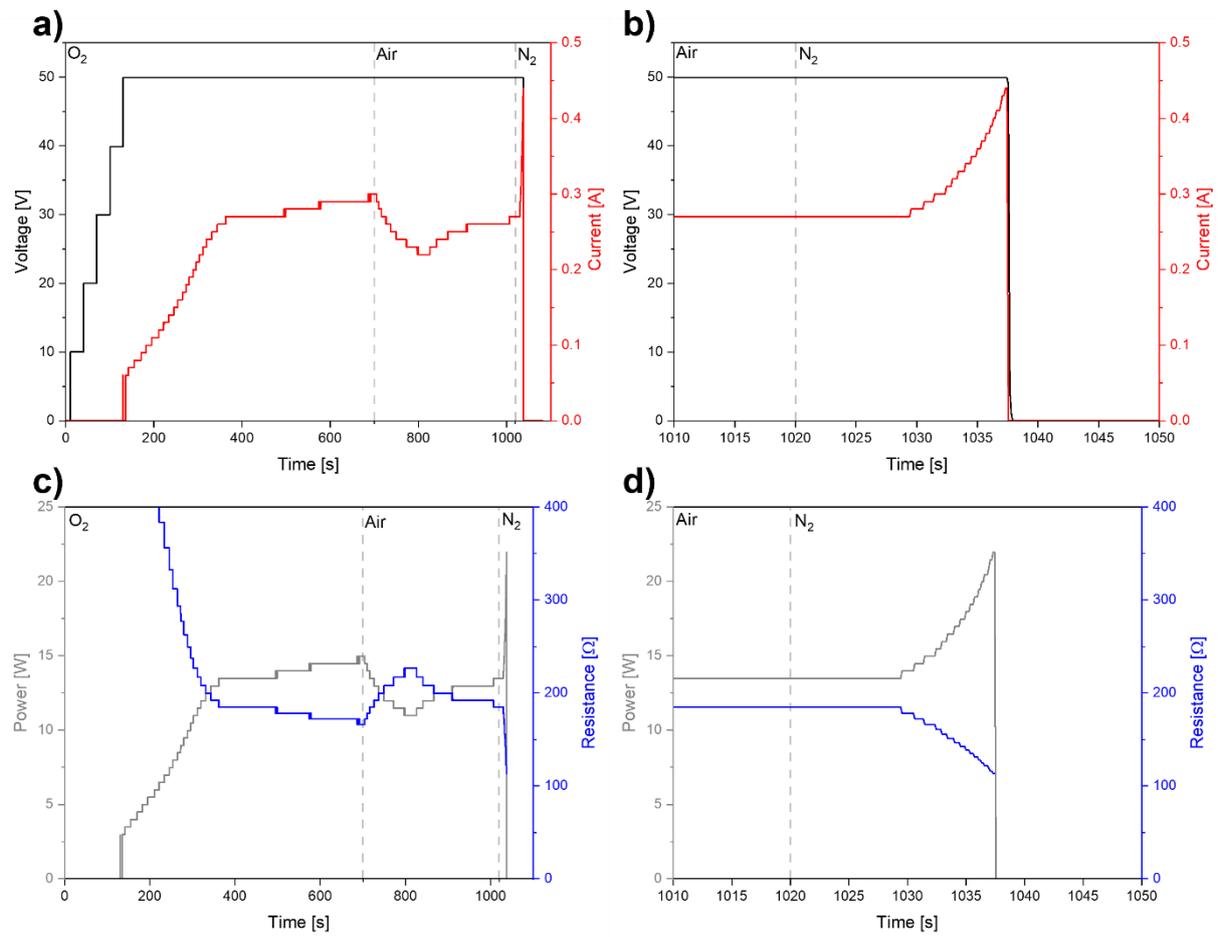

*Figure 6: Electrical parameters during atmosphere change from oxygen ($O_2$) to air to nitrogen ($N_2$) stopped after start in current increase due to electrochemical reduction.*